\documentclass[11pt,english]{article}
\usepackage{import}
\usepackage[utf8]{inputenc}
\usepackage[T1]{fontenc}
\usepackage[a4paper,       
            bindingoffset=0.2in,
            left=0.70in,
            right=0.70in,
            top=0.85in,
            bottom=0.85in,
            footskip=.25in]{geometry}
\usepackage{amsmath}
\usepackage{amsthm}    
\usepackage{amssymb}  
\usepackage{bm}       
\usepackage{moreverb}
\usepackage{graphicx}
\usepackage{accents}
\usepackage{graphicx}
\usepackage{epstopdf}
\usepackage{float}
\usepackage{footnote}
\usepackage{threeparttable}
\usepackage{breakcites}
\usepackage{blindtext}
\usepackage[parfill]{parskip}
\usepackage{multirow}
\usepackage{booktabs}
\usepackage{hyperref}
\usepackage{authblk}
\usepackage{xcolor}
\usepackage[toc,page,title]{appendix} 
\usepackage[backend=bibtex,style=numeric, sorting=none]{biblatex} 
\usepackage{breqn}
\usepackage{titlesec}
\titleformat{\paragraph}
{\normalfont\normalsize\bfseries}{\theparagraph}{1em}{}
\titlespacing*{\paragraph}
{0pt}{3.25ex plus 1ex minus .2ex}{1.5ex plus .2ex}
\DeclareUnicodeCharacter{2060}{\nolinebreak}
\makesavenoteenv{tabular}
\title{\LARGE{Meta-analysis of dichotomous and ordinal tests with an imperfect gold standard}}

\author[*,1,2]{ Enzo Cerullo }
\author[3]{ Hayley E. Jones }
\author[4]{ Olivia Carter}
\author[5]{ Terry J. Quinn }
\author[1,2]{ Nicola J. Cooper }
\author[1,2]{ Alex J. Sutton }

\affil[1]{\small{Biostatistics Research Group, Department of Health Sciences, University of Leicester, Leicester, UK}}
\affil[2]{Complex Reviews Support Unit, University of Leicester \& University of Glasgow, Glasgow, UK}
\affil[3]{Population Health Sciences, Bristol Medical School, University of Bristol, Bristol, UK}
\affil[4]{No affiliation}
\affil[5]{Institute of Cardiovascular and Medical Sciences, University of Glasgow, Glasgow, UK }

\date{}
\begin{document}

\maketitle

\section*{\large{Abstract}}

Standard methods for the meta-analysis of medical tests, without assuming a gold standard, are limited to dichotomous data. Multivariate probit models are used to analyze correlated dichotomous data, and can be extended to model ordinal data. Within the context of an imperfect gold standard, they have previously been used for the analysis of dichotomous and ordinal test data from a single study, and for the meta-analysis of dichotomous tests. However, they have not previously been used for the meta-analysis of ordinal tests.

In this paper, we developed a Bayesian multivariate probit latent class model for the simultaneous meta-analysis of ordinal and dichotomous tests without assuming a gold standard, which also allows one to obtain summary estimates of joint test accuracy. We fitted the models using the software Stan, which uses a state-of-the-art Hamiltonian Monte Carlo algorithm, and we applied the models to a dataset in which studies evaluated the accuracy of tests, and test combinations, for deep vein thrombosis. We demonstrate the issues with dichotomising ordinal test accuracy data in the presence of an imperfect gold standard, before applying and comparing several variations of our proposed model which do not require the data to be dichotomised. 

The models proposed will allow researchers to more appropriately meta-analyse ordinal and dichotomous tests without a gold standard, potentially leading to less biased estimates of test accuracy. This may lead to a better understanding of which tests, and test combinations, should be used for any given medical condition.

\section*{\large{Keywords}}

Meta-Analysis,  test accuracy, multivariate probit, latent class, imperfect gold, ordinal tests

*Corresponding Author

Email address: enzo.cerullo@bath.edu

\pagenumbering{arabic}

\newpage
\section{Introduction}
\label{section_introduction}
 
Medical tests are used to screen, monitor and diagnose medical conditions. In order to evaluate their accuracy, we can carry out test accuracy studies - studies which estimate the accuracy of a test by comparing its results to some existing test assumed to be perfect (i.e. 100\% sensitive and specific). The tests under evaluation, and those assumed to be perfect, are referred to as \textit{index} tests and \textit{reference} (or \textit{gold standard}) tests, respectively. Index tests often have a lower sensitivity and/or specificity than the gold standard; however, they may be quicker, less invasive, and/or less costly. Unfortunately, the fact that gold standard tests are often imperfect is ignored in routinely used methods \supercite{Reitsma2005, Rutter2001, Harbord2007} to meta-analyse studies of test accuracy, which can lead to misleading results\supercite{Hui1980}.

The results between tests are usually \textit{conditionally dependent} - that is, they are correlated within each disease class (diseased and non-diseased individuals). ⁠Models which account for this dependency, in addition to an imperfect gold standard have been proposed \supercite{Chu2009, Menten2013, Dendukuri2012, Sadatsafavi2010, Kang2013}. These models - which we will refer to as \textit{traditional latent class models} (TLCMs) - assume that all tests are measuring the same latent disease, and each individual is modelled as belonging in either disease class. Since they can model imperfect gold standards, they also allow one to compare the accuracy between gold standard and index tests.

Proposed models which can model an imperfect gold standard based on TLCMs have some limitations, which motivates the proposal of more flexible latent class models\supercite{Xu2009, Xu2013, Sadatsafavi2010, Uebersax, Qu1996, Qu1998, Kang2013}. For instance, multivariate probit latent class (MVP-LC) models \supercite{Xu2009, Xu2013, Sadatsafavi2010, Uebersax, Qu1996, Qu1998, Albert1993}, which are a type of regression model. Unlike TLCMs, MVP-LC models can be extended to model ordinal test accuracy data\supercite{Uebersax, Greene2012} without forcing the user to dichotomise it, whilst simultaneously modelling conditional dependence. For example, Xu et al⁠\supercite{Xu2009} presented an MVP-LC model to analyse primary studies evaluating multiple dichotomous tests without assuming a gold standard, which they later extended\supercite{Xu2013}$~$to model ordinal tests with two cutpoints. The latent trait model proposed by Qu et al\supercite{Qu1996} is a variation  of the MVP-LC model which is defined by specifying a series of univariate regressions with a common subject-specific latent variable. This model was later expanded upon by Sadatsafavi et al\supercite{Sadatsafavi2010}$~$ to the meta-analysis setting - to analyse studies evaluating up to three dichotomous tests using direct comparisons - whilst allowing the test accuracy to vary between studies. However, it cannot appropriately model ordinal tests, nor can it model between-study variation for the conditional dependence parameters.

In clinical practice, tests are rarely used in isolation. The accuracy of two or more tests used in combination is often referred to as the \textit{joint test accuracy}. Few meta-analytical methods have been proposed which can simultaneously calculate summary joint test accuracy and incorporate ordinal tests, all of which assume a perfect gold standard. For instance, Novielli et al\supercite{Novielli2013} proposed a model based on conditional probabilities, in which studies evaluated up to one ordinal test with two cutpoints and two dichotomous tests. This model can estimate summary test accuracy at each cutpoint whilst modelling conditional dependence.

To address the gaps in the literature discussed above, we developed a Bayesian model for the meta-analysis of studies evaluating both ordinal and dichotomous tests without assuming a perfect gold standard. The model also enables the estimation of summary joint test accuracy, whilst allowing the conditional dependence parameters to vary between studies. The proposed model is an extension of previous MVP-LC models which have been developed to analyse multiple tests in a single study\supercite{Xu2009, Xu2013}. In section \ref{section_case_study_dataset_motivating_example}, we describe the case study dataset which will serve to motivate our proposed model, which we will describe in section \ref{section_methods}. Then, we apply several variations of our proposed model to this dataset in section \ref{section_results_application_to_case_study}. Finally, in section \ref{section_discussion} we discuss the benefits and limitations of the model, as well as possible extensions.

\section{Motivating example} 
\label{section_case_study_dataset_motivating_example}
Deep vein thrombosis (DVT) is the formation of a blood clot in a deep (i.e. not superficial) vein. DVT can occur in the upper (proximal) or lower (distal) part of the leg, with the former more likely to be life-threatening. A potential complication of DVT occurring in up to a third of patients \supercite{stone_deep_2017} is pulmonary embolism (PE). PE occurs when a blood vessel in the lungs becomes blocked by a blood clot (formed as a result of DVT) which has migrated from the legs to the lungs.
⁠Contrast venography is generally considered to be a gold standard for DVT, as it is almost 100\% sensitive and specific\supercite{Tovey1180, Kyrle2005}. However, it is not commonly used in clinical practice because it is time consuming and invasive \supercite{Tovey1180, Kyrle2005}. 
Instead, ultrasound is often used to diagnose DVT, since it is non-invasive and cost-effective \supercite{stone_deep_2017, kearon_noninvasive_1998, ho_acr_2011, min_diagnosis_2016}. However, it is less accurate than contrast venography\supercite{Goodacre2005, DiNisio2010} for both distal and proximal DVT, with its sensitivity being lower for distal DVT \supercite{Goodacre2005}. Furthermore, although ultrasound is known to have a very high specificity, it is still nonetheless imperfect\supercite{Goodacre2005, DiNisio2010}.
A commonly used\supercite{stone_deep_2017} screening tool for DVT is a questionnaire called the Wells score\supercite{wells_value_1997}, which groups patients into one of three risk categories - 'low', 'intermediate', or 'high'.
Another DVT test is the D-Dimer assay: a blood test measuring the amount of a protein fragment called D-Dimer, higher concentrations of which are indicative of DVT. Despite being considered to be generally more accurate than the Wells\supercite{khan_venous_2021}, the D-Dimer assay is intended to be used for screening as opposed to diagnosis\supercite{khan_venous_2021}, since a number of other conditions can elevate serum D-dimer concentrations\supercite{stone_deep_2017}. 

Investigating the joint test accuracy of the aforementioned tests for DVT is important for a variety of reasons. 
The Wells and D-Dimer are both relatively cheap, quick and non-invasive to carry out, particularly the Wells test.  A combined screening approach utilising the Wells and D-Dimer may be more cost-effective and reduce test burden for patients compared to using either alone. Furthermore, despite the fact that neither the Wells nor the D-Dimer alone are generally considered to be diagnostic tools for DVT, they may have diagnostic potential when combined\supercite{Novielli2013, Goodacre2005, Novielli_2013_econimic}.  
An example of a potential screening strategy is to use the Wells prior to the D-dimer in the diagnostic pathway as a pre-screening tool to rule out individuals at low risk for DVT. Following this, individuals who scored as intermediate or high risk are subsequently screened using the D-Dimer assay, and only patients who also test positive on the D-Dimer undertake ultrasound. 
Another potential strategy is to refer patients scoring as high risk on the Wells score directly to ultrasound.
Both of the aforementioned joint testing strategies are examples of \textit{‘believe the negatives’} (BTN) strategies \supercite{Novielli2013}. This is a testing strategy where only those patients who test positive on an initial test go on to receive a second test, then only individuals who also test positive on the second test are considered positive. 
Conversely, \textit{‘Believe the positives’} (BTP) is a testing strategy where only those patients who test negative on the first test go on to receive a second test, with only those patients who also test negative on this test being considered negative.
Joint testing strategies are important across clinical areas besides DVT, for example for depression screening and for COVID-19 - see discussion section \ref{section_discussion_summary} for more details. 

Novielli et al⁠\supercite{Novielli2013} proposed a statistical model in order to conduct a meta-analysis of studies investigating the D-dimer, Wells score and ultrasound for DVT. The proposed model allowed them to model the Wells score without dichotomising the data whilst modelling the conditional dependence between tests, enabling them to estimate summary-level joint test accuracy. However, their model assumes that ultrasound is a perfect gold standard, which could have led to biased estimates of the performance of other tests under evaluation. Novielli et al\supercite{Novielli2013} carried out several analyses based on different datasets – for instance, one based on the 11 studies which directly compared the D-dimer, Wells’ score via the gold standard (ultrasound), and another which also included studies which only analysed one of Wells or D-dimer tests, and utilised indirect comparisons. 
In section \ref{section_results_application_to_case_study} of this paper, we re-analyse the direct comparisons data (see table \ref{table_data}) from Novielli et al \supercite{Novielli2013} without assuming a perfect gold standard, using a variety of models we propose in section \ref{section_methods}; namely, models which dichotomised the Wells score and those which modelled it as an ordinal test, those which assumed conditional independence and dependence between tests, as well as models which assumed ultrasonography was perfect or imperfect. This dataset consisted of 11 studies, with a total of 4096 individuals and 12,288 observations, with all 11 studies evaluating all three tests.

\begin{table}[H]
\centering
\begin{threeparttable}
\caption{Sample of case study dataset}
\begin{tabular}{@{}lllllllllllll@{}}
\toprule
\multirow{4}{*}{Study} & \multicolumn{6}{l|}{Ultrasound -'ve}                                                                                                                & \multicolumn{6}{l|}{Ultrasound +'ve}                                                                                                               \\ \cmidrule(l){2-13} 
                       & \multicolumn{3}{l|}{D-Dimer -'ve}                                        & \multicolumn{3}{l|}{D-Dimer +'ve}                                        & \multicolumn{3}{l|}{D-Dimer -'ve}                                        & \multicolumn{3}{l|}{D-Dimer +'ve}                                       \\ \cmidrule(l){2-13} 
                       & \multicolumn{6}{l|}{Wells score\tnote{1}}                                                                                                                    & \multicolumn{6}{l|}{Wells score\tnote{1}}                                                                                                                   \\ \cmidrule(l){2-13} 
                       & \multicolumn{1}{l|}{L} & \multicolumn{1}{l|}{M} & \multicolumn{1}{l|}{H} & \multicolumn{1}{l|}{L} & \multicolumn{1}{l|}{M} & \multicolumn{1}{l|}{H} & \multicolumn{1}{l|}{L} & \multicolumn{1}{l|}{M} & \multicolumn{1}{l|}{H} & \multicolumn{1}{l|}{L} & \multicolumn{1}{l|}{M} & H                     \\ \midrule
1                      & 32                     & 20                     & 5                      & 8                      & 18                     & 2                      & 0                      & 0                      & 2                      & 1                      & 6                      & 8                     \\
$\vdots $  &  $\vdots$  & $\vdots$  & $\vdots$  & $\vdots$  & 
 $\vdots$  & $\vdots$  & $\vdots$  & $\vdots$  & $\vdots$  & $\vdots$  & $\vdots$  & $\vdots$ \\
11                     & 243                    & 16                     & 3                      & 233                    & 104                    & 29                     & 1                      & 0                      & 0                      & 28                     & 117                    & 109                   \\ \bottomrule
\label{table_data}
\end{tabular}

   \begin{tablenotes}
    \item Note: All test results are modelled at the individual level. We show the aggregate data in this table for ease of presentation.
    \item[1] The Wells score is classified as L = Low, M = Moderate, H = High
   \end{tablenotes}
     \end{threeparttable}
\end{table}

\section{Methods} 
\label{section_methods}
Before describing our proposed Bayesian MVP-LC model, we will first define some terminology and notation in section \ref{section_methods_terminology_and_notation}. For a formal model specification, please refer to the full technical model specification (supplementary material 1). 

\subsection{Terminology \& notation}
\label{section_methods_terminology_and_notation}
The model is for a meta-analysis dataset with a total of $S$ studies, with each study having a total of $N_s$ individuals, where $s$ is an index for study - so $s$ can be used to denote anything between the first ($s=1$) and the last ($s=S$) study. Each study is assessing the same number of tests, $T$. Will we use $t$ as an index for test, which can be between $1$ and $T$, and $n$ as an index for individual, which can be between $1$ and $N_s$ for study $s$. 

For the $n$th individual from study $s$, we will denote the vector of observed test responses as $\mathbf{y}_{s,n}  = (y_{s,n,1}, \hdots , y_{s,n,T})'$. Each test is either dichotomous or ordinal. For dichotomous tests, each observed test response, $y_{s,n,t}$, is coded as 0 and 1 for negative and positive results, respectively. For ordinal tests, each test $t$ has $K_t$ categories (hence $K_t - 1$ cutpoints). We will use $k$ as an index to refer to any given cutpoint, which can be between $1$ and $K_t - 1$. Ordinal test responses are coded according to the category that the individuals' test result falls in: in other words, $y_{s,n,t} = k$ if the test result falls in the $k$th category for test $t$. Since we are assuming an imperfect gold standard, the true disease status of each individual, $d_{s,n}$ is not defined by the results of the gold standard. Instead, it is modelled as an unknown (i.e. \textit{latent}) variable, and belongs to one of two classes - 'diseased' ($d_{s,n} = 1$) or 'non-diseased' ($d_{s,n} = 0$).

\subsection{Within-study model}
\label{section_methods_within_study_model}

We will now define the MVP-LC model within each study. For dichotomous data, MVP-LC models work by transforming the observed, discrete test result data into a \textit{continuous} variable - a statistical technique known as "data augmentation" \supercite{Albert1993}. This augmented continuous data, which we will denote as $\mathbf{Z}_{s,n}$, is an unobserved latent variable, similarly to the disease status, $d_{s,n}$. This data augmentation process allows us to assume that, conditional on $d_{s,n}$, the unobserved test accuracy data,  $\mathbf{Z}_{s,n}$, can be modelled by a multivariate normal (MVN) distribution with mean vector $\boldsymbol\nu_{s}^{[d]}$ and variance-covariance matrix $\boldsymbol\Psi^{[d]}_{s} $. 

More specifically, $\mathbf{Z}_{s,n} \sim \text{MVN} ( \boldsymbol\nu_{s}^{[d]} ,  \boldsymbol\Psi^{[d]}_{s} )$, where:

\begin{equation}
\begin{aligned}
\mathbf{Z}_{s,n} =  
 \begin{pmatrix}
    Z_{s,n,1}   \\  \vdots   \\  Z_{s,n,T}     \\ 
\end{pmatrix} 
,
\boldsymbol\nu_{s}^{[d]} = 
\begin{pmatrix}
    \nu^{[d]}_{s,1}   \\    \vdots   \\    \nu^{[d]}_{s,T}     \\ 
\end{pmatrix}
,
\boldsymbol\Psi^{[d]}_{s} =  
\begin{pmatrix}
 (\tau^{[d]}_{s,1})^2 & \cdots & \dot 
 \epsilon^{[d]}_{s,1,T} \cdot \tau^{[d]}_{s,1} \cdot \tau^{[d]}_{s,T} \\ \vdots &  \ddots & \vdots \\ \dot 
 \epsilon^{[d]}_{s,T,1} \cdot \tau^{[d]}_{s,T} \cdot \tau^{[d]}_{s,1} & \cdots & 
 (\tau^{[d]}_{s,T})^2 
\end{pmatrix}
\end{aligned}
\label{eq_1_within_study}
\end{equation}

Where $\nu_{s,t}^{[d]}$ and $\tau^{[d]}_{s,t}$, denote the study-specific means and standard deviations, respectively. 
Each $ \epsilon^{[d]}_{s,1,t}$ denotes the study-specific correlations between each test-pair (or 'test-pair' - denoted as '$t$ and $t'$') for the augmented data ($Z_{s,n,t}$ and $Z_{s,n,t'}$) - the \textit{polychoric correlation} \supercite{Greene2012, Ekstrom2011} - which is not the same as the correlation between the observed data $y_{s,n,t}$ and $ y_{s,n,t'}$. 
Each $ \epsilon^{[d]}_{s,1,t}$ models the conditional dependence between each test-pair. 
However, we can assume \textit{conditional independence} by setting $\epsilon^{[d]}_{s,1,t} = 0$ and $\epsilon^{[d]}_{s,1,t'} = 0$. 
We must to ensure that the number of parameters being estimated from our model is not greater than what is possible for the given dataset; otherwise it may be \textit{non-identifiable} - which means that the model will give misleading results. For example, it may estimate the sensitivity for a test to be equal to both \textit{both} 0.20 and 0.80.
To ensure that our model is identifiable \supercite{Uebersax, Greene2012}, as in Xu et al  \supercite{Xu2013}, we set each $ \tau^{[d]}_{s,t} = 1 $ (i.e., set all $\boldsymbol\Psi^{[d]}_{s} $ to be correlation matrices).
Please see supplementary material 6. 

For dichotomous tests, the augmented data ($Z_{s,n,t}$) will be less than 0 for negative results ($y_{s,n,t} = 0$) or greater than 0 for positive results ($y_{s,n,t} = 1$), and the measures of test accuracy for a given study $s$ are given by,
\begin{equation}
\begin{aligned}
Se_{s,t}  & =     \Phi( \nu_{s,t}^{[1]}  )  \\ 
Sp_{s,t}  & = 1 - \Phi( \nu_{s,t}^{[0]}  )  
\end{aligned}
\label{eq_2_study_specific_accuracy_estimates_dichotomous}
\end{equation}
Where $\Phi(\cdot)$ denotes the cumulative density function (CDF) of the standard normal distribution - that is, a normal distribution with mean 0 and standard deviation 1. For ordinal tests, the augmented data ($Z_{s,n,t}$) will belong to an interval defined by strictly increasing latent cutpoint parameters ( $\{C^{[d]}_{1,s,t}, ..., C^{[d]}_{K_{t}-1,s,t}\}$, where $C^{[d]}_{k-1,s,t} <  C^{[d]}_{k,s,t}$, and $k$ between $2$ and $K_{t}-1$). This interval will depend on the observed test result as follows - 
if the test result is below the first cutpoint (i.e. in the first category), then the augmented data will be less than the first cutpoint parameter; 
if it is above the last cutpoint (i.e., in the last category), then the augmented data will be greater than the last cutpoint parameter; 
otherwise, if the test result falls between two cutpoints (i.e., the test result belongs to any other category), then the augmented data will fall between the corresponding cutpoint parameters. 
The measures of test accuracy are given by, 
\begin{equation}
\begin{aligned}
Se_{s,t,k} &  =  1 - \Phi( \nu_{s,t}^{[1]}  - C^{[1]}_{k,s,t} ) \\
Sp_{s,t,k}  & =  \Phi( \nu_{s,t}^{[0]}  - C^{[0]}_{k,s,t} ) 
\end{aligned}
\label{eq_3_study_specific_accuracy_estimates_ordinal}
\end{equation}
\subsection{Between-study model}
\label{section_methods_between_study_model}
Now we will explain how we will model the variation in test accuracy between studies - called the \textit{between-study heterogeneity}, as well as the correlation between the sensitivities and specificity between studies - called the \textit{between-study correlation}. It is important to bear in mind the distinction from the within-study correlations (defined in section \ref{section_methods_within_study_model}), which model the conditional dependence between tests. For each test $t$, we will assume that the study-specific means ($\nu_{s,t}^{[d]}$ - defined in equation \ref{eq_1_within_study} in section \ref{section_methods_within_study_model}) arise from a bivariate normal (BVM) distribution with means $\mu_{t}^{[d]}$, between-study standard deviations $\sigma_{t}^{[d]}$, and between-study correlations $ \rho_{t} $. 

More specifically, $ \boldsymbol\nu_{s,t} \sim \text{BVN}(\boldsymbol\mu_{t}, \boldsymbol\Sigma_{t}) $, where,

\begin{equation}
        \begin{aligned}
        \boldsymbol\mu_{t} = 
                \begin{pmatrix}
                \mu_{t}^{[1]} \\ 
                \mu_{t}^{[0]} 
                \end{pmatrix} 
        ,
         \boldsymbol\Sigma_{t} = 
                 \begin{pmatrix} 
                 \left(\sigma_{t}^{[1]}\right)^{2} &
                 \rho_{t}  \cdot \sigma_{t}^{[1]}  \cdot \sigma_{t}^{[0]} \\
                 \rho_{t}  \cdot \sigma_{t}^{[1]}  \cdot \sigma_{t}^{[0]} & 
                 \left(\sigma_{t}^{[0]}\right)^{2}
                 \end{pmatrix} 
        \end{aligned}
\label{eq_4_MVN_between_study}
\end{equation}
The model described in equation \ref{eq_4_MVN_between_study} is known as a \textit{partial pooling}  model (using the terminology from Gelman \& Hill\supercite{Gelman2006} - otherwise known as  \textit{'random-effects'}). These models allow the study-specific accuracy parameters across studies to inform one another, without assuming full homogeneity like a full pooling (i.e., “fixed-effects) would – which would allow no between-study variation in the means $\nu_{s,t}^{[d]}$. The disease prevalence's in each study, $p_s$, are modelled independently of each other, known as a \textit{no pooling} model. 
There are several differences between partial pooling and no pooling models \supercite{Betancourt2020_hierarchical}. For example, the former uses less parameters than no pooling, which means that there is less likelihood of encountering parameter identifiability issues. An advantage of our partial pooling model is that allows us to summarise the results using the parameters which are shared across studies (see section \ref{section_methods_between_study_model}), allowing us to more easily summarise test accuracy as well as the heterogeneity in accuracy between studies and correlation between sensitivities and specificities.
We can incorporate meta-regression covariates into the model by extending the partial pooling model defined in equation \ref{eq_4_MVN_between_study} above - see supplementary material 1, meta-regression section (section 1.2.1) for details. We can assume that a given test is a perfect gold standard by setting $\mu_{t}^{[0]} = -5 $ and $\mu_{t}^{[1]} = 5 $, which correspond to approximately 100\% sensitivity and specificity, respectively, and, by assuming a complete pooling model (i.e. setting $\sigma_{t}^{[d]} = 0 $ ).
\subsubsection{Within-study correlations}
\label{section_methods_between_study_model_for_within_study_correlations}
We will model the within-study correlation matrices ($\boldsymbol{\Psi}_{s}^{[d]}$) defined in equation (\ref{eq_1_within_study}) in section \ref{section_methods_within_study_model}  using a partial pooling model. As suggested by Goodrich \supercite{Goodrich2016}, this can be achieved by specifying each $\boldsymbol{\Psi}_{s}^{[d]}$ as a weighted linear combination of a global 'average' correlation matrix across studies (${\boldsymbol{\Psi}}_G^{[d]} $), and a matrix of study-level deviations from this global matrix ($ {\boldsymbol{\Psi}}_s^{{[d]}^{\Delta}} $), with weight $ \beta^{[d]} $ which is between 0 and 1. More specifically, 
$ \boldsymbol{\Psi}_{s}^{[d]} = \left(1 - \beta^{[d]} \right) \cdot {\boldsymbol{\Psi}}_G^{[d]}  + \beta^{[d]} \cdot {\boldsymbol{\Psi}}_s^{{[d]}^{\Delta}} $. Note that we can also model the conditional dependence between only certain pairs of tests by setting the relevant terms for the other tests in  ${\boldsymbol{\Psi}}_G^{[d]} $ and $ {\boldsymbol{\Psi}}_s^{{[d]}^{\Delta}} $ to zero. 
\subsubsection{Cutpoints}
\label{section_methods_between_study_model_cutpoints}
The cutpoint parameters can be modelled using an induced Dirichlet model, an approach which has been proposed by Betancourt \supercite{Betancourt2019_ordinal} By taking advantage of the properties of a type of statistical distribution called a Dirichlet distribution, this model is able to map the latent cutpoint parameters in each study ($\{ C^{[d]}_{1,s,t}, \hdots, C^{[d]}_{K_t-1,s,t} \}$) defined in section \ref{section_methods_within_study_model} to a simplex (i.e., a vector whose elements sum to 1) of ordinal probabilities ($ P^{[d]}_{1,s,t}, \hdots, P^{[d]}_{K_t,s,t} $). Each probability $P^{[d]}_{k,s,t}$ corresponds to the probability that an individuals' test result for test $t$ falls in category $k$ for study $s$. In this paper, we used a partial pooling model for the cutpoints accross studies, enabling us to model the between-study heterogeneity in the cutpoints. We can also obtain 'average' cutpoints, ($\mathbf{C}^{[d]}_{t}$) by using the posterior distribution of the induced Dirichlet partial pooling model, enabling us to obtain summary accuracy measures for ordinal tests. For the full details of this model, see supplementary material 1 and supplementary material 4.
\subsubsection{Test accuracy summaries}
\label{section_methods_between_study_model_test_accuracy_summaries}
For dichotomous tests, the summary sensitivity and specificity estimates for test $t$ are given by evaluating equation (\ref{eq_2_study_specific_accuracy_estimates_dichotomous}) at the means of the between-study model (see equation (\ref{eq_1_within_study})). More specifically, 
$Se_{G,t}  = \Phi( \mu_{t}^{[1]}) $
, and 
$Sp_{G,t}  = 1 - \Phi( \mu_{t}^{[0]}  )$. Similarly, for ordinal tests, the summary measures for test $t$ at cutpoint $k$ are given by evaluating equation (\ref{eq_3_study_specific_accuracy_estimates_ordinal}) evaluated at the means of the partial pooling model (see equation (\ref{eq_1_within_study})), and at the global (summary) cutpoints ($C^{[1]}_{k,t}$). That is, 
$ Se_{G,t,k} =  1 -\Phi( C^{[1]}_{k,t} - \mu^{[1]}_{t}) $
, and 
$ Sp_{G,t,k}  =     \Phi( C^{[0]}_{k,t} - \mu^{[0]}_{t}) $. 
We can generate predictions for a 'new' ($\textit{S+1}$)-th study by simulating a draw (at each iteration of the parameter sampler) from the posterior predictive distributions of the between-study normal hierarchical model, (see (\ref{eq_4_MVN_between_study})), 
$ \boldsymbol\nu_{S+1,t} $, 
and a new vector of cutpoints from the between-study cutpoint model
(for more details, see section 1.2.4 in supplementary material 1).
$ \mathbf{C}^{[d]}_{S+1, t})  $.
Then, the predicted sensitivities and specificities for an ($S+1$)-th study are given by
$ Se_{S+1,t}  =   \Phi( \nu_{S+1, t}^{[1]}  ) $,
$ Sp_{S+1,t}  =  1 - \Phi( \nu_{S+1, t}^{[0]}   )  $ for dichotomous tests, and 
$ Se_{S+1,t,k} =  1 -\Phi( C^{[1]}_{S+1,k,t} - \nu^{[1]}_{S+1, t})  $,
$ Sp_{S+1,t,k}  =  \Phi( C^{[0]}_{S+1, k,t} - \nu^{[0]}_{S+1, t}) $ for ordinal tests. 
\subsubsection{Joint test accuracy summaries}
\label{section_methods_between_study_model_joint_test_accuracy}
The summary estimates for the joint test accuracy of tests $t$ and $t'$ at cutpoints $k$ and $k'$ are given by: 
\begin{equation}
\begin{aligned}
Se^{BTN}_{G, tt', kk'} & = Se_{G,t,k}*Se_{G,t',k'} + cov_{G, tt', kk'}^{[1]}  \\
Sp^{BTN}_{G, tt', kk'} & = 1 - ( (1- Sp_{G,t,k})*(1 - Sp_{G,t',k'} ) + cov_{G, tt', kk'}^{[0]} ) \\
Se^{BTP}_{G, tt', kk'} & = 1 - ( (1 - Se_{G,t,k})*(1 - Se_{G,t',k'}) + cov_{G, tt', kk'}^{[1]} ) \\
Sp^{BTP}_{G, tt', kk'} & = Sp_{G,t,k}*Sp_{G,t',k'} + cov_{G, tt', kk'}^{[0]}  \\
\end{aligned}
\label{eq_5_joint_test_accuracy}
\end{equation}
With BTN and BTP as defined in section \ref{section_case_study_dataset_motivating_example}. 
Note that for ordinal tests, the order of the tests can affect joint test accuracy estimates  \supercite{Novielli_2013_econimic}. However, for dichotomous tests it does not\supercite{Novielli_2013_econimic}, although this is often still important for clinical practice. For example, the first test may be cheaper to carry out. 
The parameter $cov_{G, tt', kk'}^{[d]}$ is the global conditional covariance between all possible test-pairs.
Obtaining these covariances requires us to calculate the global conditional correlations between each test-pair, which we will denote as $ \rho_{G,tt',kk'}^{[0]}$; 
this assumes that the test results are the same form as the observed data. 
However, our model is parameterised in terms of the polychoric correlations ($ \dot\epsilon_{G,tt',kk'}^{[d]}$ - see equation (equation \ref{eq_1_within_study}) in section \ref{section_methods_within_study_model}).
Therefore, in order to be able to estimate the joint test accuracy estimates in equation \ref{eq_5_joint_test_accuracy}, we will need to convert from $ \dot\epsilon_{G,tt',kk'}^{[d]}$ to $ \rho_{G,tt',kk'}^{[d]}$. 
For details on how this is achieved, please refer to section 1.2.4 of supplementary material 1.
\subsection{Assessing model fit \& model comparison }
\label{section_methods_model_fit_and_comparison}

For our MVP-LC model, We can check how well our model predicts the data by using a technique called \textit{posterior predictive checking} - where we generate data from our model, and compare it to the observed data. For example, we can plot the model-predicted test results against the observed test results for each test-pair \supercite{Chu2009, Hui1980, Menten2013}. We also assessed model fit by plotting the model-predicted within-study correlations against the observed within-study correlations, using the correlation residual plot proposed by Qu et al\supercite{Qu1996}. For model comparison, we used leave-one-out (LOO) cross-validation \supercite{Vehtari2017} - an iterative procedure which removes part of the data and re-fits the model, and sees how well the model predicts the missing data. For more details on model comparison and posterior predictive checking, including relevant formulae, please refer to 
section 1.3 in supplementary material 1. 

\subsection{Model implementation}
\label{section_methods_model_implemention}
We implemented the models in R \supercite{R_software_ref} using the probabilistic programming language Stan \supercite{Carpenter2017, Betancourt2017_hmc_intro} via the R package CmDStanR\supercite{CmDStanR} using a PC with 32GB of RAM and an AMD Ryzen 3900X 12-core CPU with Linux Mint OS. To code the model in Stan, we extended the code for a standard binary multivariate probit model \supercite{Goodrich2016}⁠$~$. This is described in detail in Goodrich 2017\supercite{Goodrich2017} and is summarised in supplementary material 5. We implemented the between-study partial pooling model for the within-study correlations described in section \ref{section_methods_between_study_model_for_within_study_correlations} in Stan by using the function provided by Stephen Martin and Ben Goodrich\supercite{Martin2018}. For the cutpoint between-study model, we used Betancourt's induced Dirichlet model\supercite{Betancourt2019_ordinal} described in section \ref{section_methods_between_study_model_cutpoints}; this is described in more detail in supplementary material 4, and this was implemented using code by Betancourt \supercite{Betancourt2019_ordinal}. 

We ran all models using 4 chains until the split R-hat statistic was less than 1.05 for all parameters and the number of effective samples was satisfactory for all parameters \supercite{stan}. We only reported results when we obtained no warnings for divergent transitions or energy fraction of missing information (E-FMI), important diagnostics for geometric ergodicity \supercite{Betancourt2017_hmc_intro}. We used the CmDStanR diagnostic utility to check all of the aformentioned model diagnostics \supercite{CmDStanR}. We also inspected trace plots and plotted the posterior distributions to check they were not bimodal. Rather than using $\Phi(\cdot)$, which is prone to numerical instability, we can use the closely resembling logistic function, $\Phi'\left(x\right) = \frac{1}{1+ e^{-1.702 \cdot x}} $, which has an absolute maximum deviation from $\Phi(\cdot)$ of 0.0095. This is the same probit approximation used for the meta-analysis of dichotomous tuberculosis tests using latent trait models in Sadatsafavi et al\supercite{Sadatsafavi2010}. The data, Stan model code, and R code to reproduce the results and figures for the case study application in section \ref{section_results_application_to_case_study} is provided at \href{https://github.com/CerulloE1996/dta-ma-mvp-1}.
\section{Application to case study}
\label{section_results_application_to_case_study}

Since our model is Bayesian, we must formulate a \textit{prior model} - that is, specify prior distributions for the model parameters defined in section \ref{section_methods}. We describe this prior model in \ref{section_results_priors}. 
When faced with the task of analysing a dataset with an imperfect gold standard which contains test accuracy data from an ordinal test, in order to be able to apply proposed methods for meta-analysis without assuming a gold standard \supercite{Chu2009, Dendukuri2012}, one must first dichotomise the data at each cutpoint and conduct a series of stratified analyses. We applied a priori dichotomisation technique using our proposed MVP-model in section \ref{section_results_a_priori_dichot}.
Finally, in section \ref{section_results_modelling_wells_as_ordinal}, we applied the models proposed in section \ref{section_methods}, but without dichotomising the Well's score. 

In section \ref{section_results_priors}, we will index the gold standard (ultrasound), the D-Dimer, and the Wells' score by $t=1, t=2$, and $t=3$, respectively. In sections \ref{section_results_a_priori_dichot} and \ref{section_results_modelling_wells_as_ordinal} we will denote summary estimates as "X [Y, Z]", where X is the posterior median and [Y, Z] is the 95\% posterior interval. 

\subsection{Prior distributions}
\label{section_results_priors}

For the summary-level accuracy parameter for the gold standard test (ultrasound - i.e. $\mu_{t=1}^{[d]}$), we constructed informative priors using subject-matter knowledge for ultrasound, based on meta-analyses from the literature \supercite{Goodacre2005, DiNisio2010} (see supplementary material 2 for more details). These priors correspond to 95\% prior intervals of $(0.49, 0.94)$ and $(0.82, 0.99)$ for the sensitivity and specificity, respectively. 
On the other hand, for the summary-level accuracy parameters for the D-Dimer and the Wells score (i.e., $\mu_{t=2}^{[d]}$ and $\mu_{t=3}^{[d]}$), we specified priors conveying very little information - equivalent to assuming a 95\% prior interval of $(0.04, 0.96)$ for the sensitivities and specificities. 

For the between study deviation parameters for all three tests (i.e., for $ \sigma_{t}^{[d]}$ for $t = \{1, 2, 3\}$ - see equation (\ref{eq_4_MVN_between_study}) in section \ref{section_methods_between_study_model}), we used weakly informative priors corresponding to a 95\% prior interval of $(0.02, 1.09)$. The priors are weakly informative since they weakly pull the study-specific sensitivities and specificities towards each other, whilst allowing for large between-study heterogeneity if the data demands. For example, if $0.8$ is the value found for the summary sensitivity, and the data suggests a standard deviation equal to 2 (corresponding to a high degree of between-study heterogeneity), then these priors would allow the study-specific sensitivities and specificities to be in the interval $(0.44, 0.97)$ with 95\% probability. 
We also used weak priors for the between-study correlation parameters for all tests (i.e., $\rho_{t}$ for $ t = \{1, 2, 3\}$ - see equation (\ref{eq_4_MVN_between_study} in section \ref{section_methods_between_study_model}), corresponding 95\% prior probability interval of (-0.82, 0.82). 
Finally, for conditional dependence models, for the within-study correlation parameters (see section \ref{section_methods_within_study_model} and equation \ref{eq_1_within_study}), we used priors which correspond to 95\% prior intervals of $(-0.65, 0.65)$ for both the global 'average' correlation matrices ($\boldsymbol\Omega_{G}^{[d]}$) and the study-specific deviation matrices ($ \boldsymbol\Omega_{s}^{{[d]}^{\Delta}}$), respectively. These are weakly informative and allow a moderately large between-study deviation in the strength of the conditional dependence between tests. For more detail on these prior distributions, please see supplementary material 2.

\subsection{The pitfalls of a priori dichotomisation in the presence of an imperfect gold standard}
\label{section_results_a_priori_dichot}
We consider two dichotomisations of the Wells score. For the first, we dichotomised the Wells' score by grouping together those patients who obtain a score of 'low' or 'moderate' as a negative result and those who scored 'high' as positive.
On the other hand, for the second dichotomisation. we grouped together patients who scored 'moderate' or 'high' and considered this as a positive result, and those who scored 'low' as a negative result. 
We will refer to the former dichtomisation as "low + moderate vs high" and the latter as "low vs moderate + high". We applied this technique to this dataset, , to allow comparison with our "full" model, using the models proposed in section \ref{section_methods}, fitting both conditional independence (CI) and dependence (CD) models, the results of which are shown in  in section 

\begin{figure}[H]
    \centering
    \includegraphics[width=15cm]{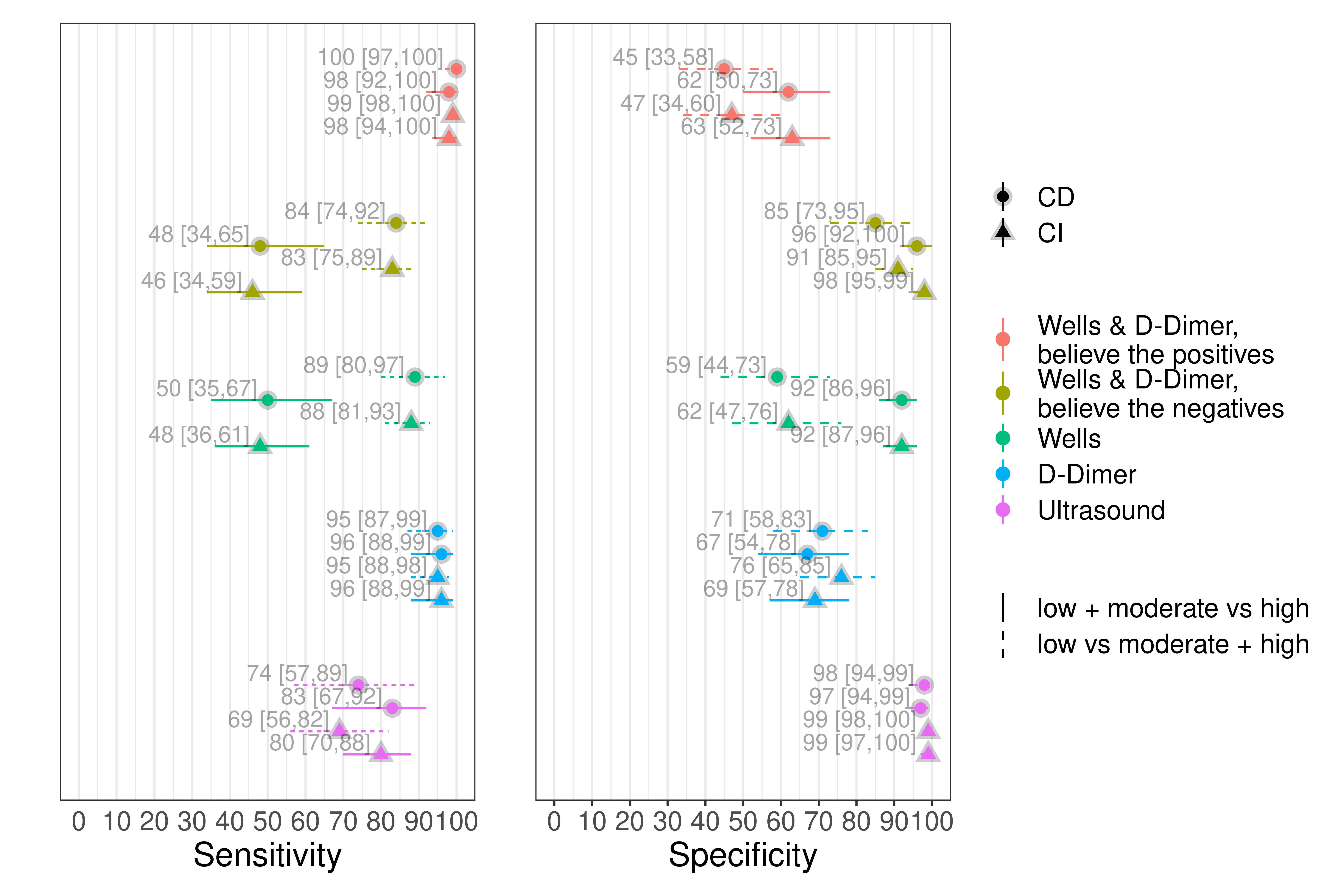}
    \caption{Posterior medians and 95\% posterior intervals for models dichotomising the Well's score. 
    Note: CD = Conditional Dependence; CI = Conditional Dependence}
\label{fig1}
\end{figure}

\begin{figure}[H]
    \centering
    \includegraphics[width=12cm]{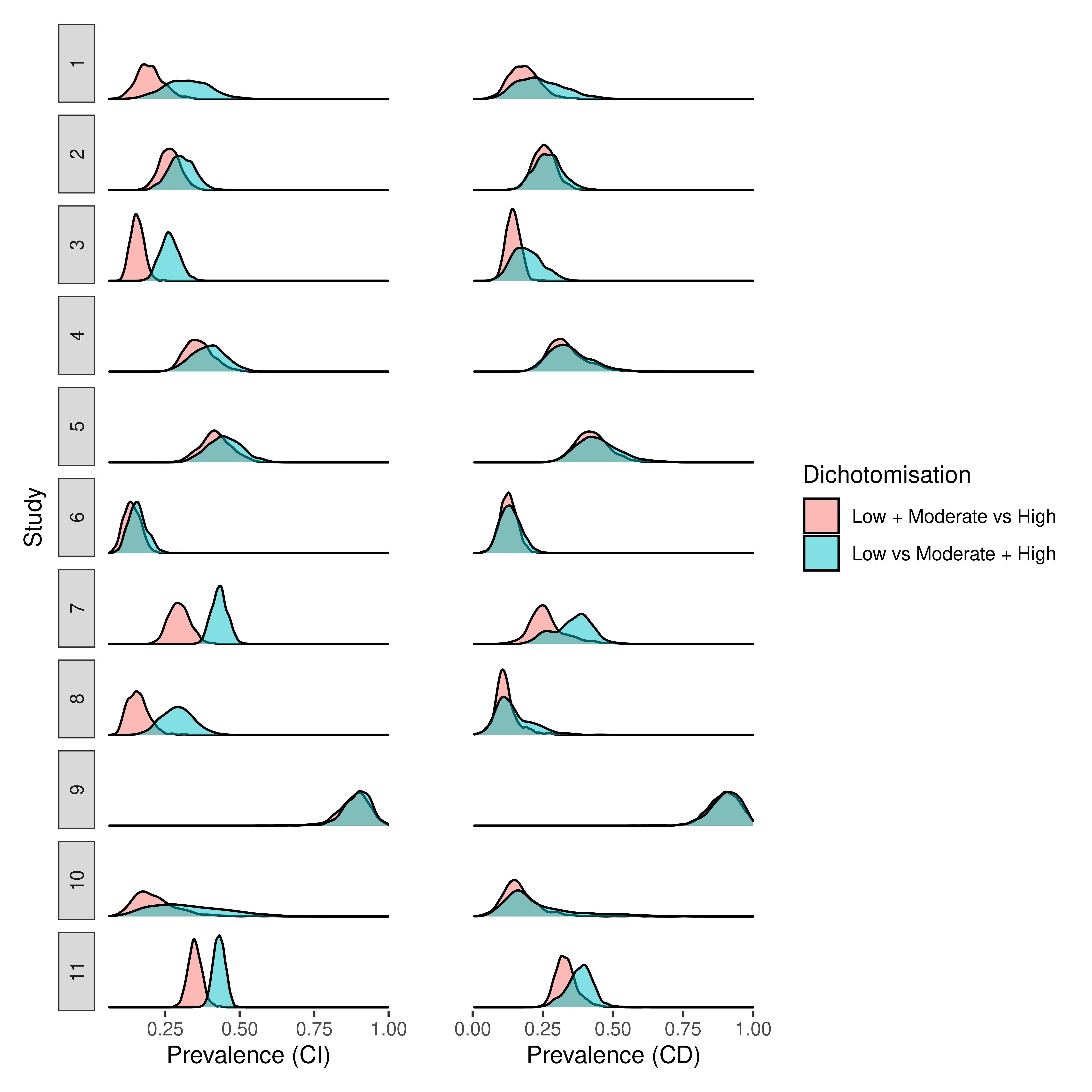}
    \caption{Posterior density plots for disease prevalence parameters.
    Note: CD = Conditional Dependence; CI = Conditional Dependence}
\label{fig2}
\end{figure}

When assuming conditional independence between all three tests, we see that (figure \ref{fig1}) some of the estimates of the accuracy of the other two tests change substantially depending on whether we dichotomise the Wells score as low+moderate vs high, or as low vs moderate+high. For the former dichotomisation, the sensitivity of ultrasound was estimated as 0.80 [0.70, 0.88] whereas for the latter it was 0.69 [0.56, 0.82]. The specificity of ultrasound and the sensitivity of the D-Dimer were similar between both dichotomisations. However, there was a notable difference in the specificities of the D-Dimer test, where we obtained specificities of 0.69 [0.57, 0.78] and 0.76 [0.65, 0.85] for the low+moderate vs high and low vs moderate+high dichotomisations, respectively. 

The differences in the results were similar when modelling conditional dependence between the three tests (see figure \ref{fig1}). In the low+moderate vs high dichotomisation, for the ultrasound sensitivity we obtained 0.83 [0.67, 0.92] and for the low vs moderate + high dichotomisation 0.74 [0.57, 0.89]. For the D-Dimer specificities, we obtained 0.67 [0.54, 0.78] and 0.71 [0.58, 0.83] for the low+moderate vs high and low vs moderate+high dichotomisations, respectively. As with conditional independence, the specificity of the ultrasound and the sensitivity of the D-Dimer were similar between the two dichotomisations. We can also see the estimates of disease prevalence increase for most studies for the low vs moderate + high dichotomisation relative to the low + moderate vs high dichotomisation, for both conditional independence (left panel of figure \ref{fig2}) and dependence models (right panel of figure \ref{fig2}). 

Overall, regardless of whether we assume conditional independence or dependence, some of the accuracy estimates change notably depending on how we dichotomise the Wells score. This is not surprising, since imperfect gold standard models based on latent class analysis utilise the full distribution of test responses from all tests to estimate accuracy and disease prevalence \supercite{Hui1980}. This simple example demonstrates the importance of modelling all the available data for ordinal non-dichotomous tests, such as the Wells score, in the presence of an imperfect gold standard, as opposed to simply conducting multiple stratified analyses at each cutpoint of the ordinal test using simpler methods. This observation serves to motivate the implementation of ordinal regression into the models to appropriately model the ordinal nature of the Wells score. 
\subsection{Modelling the Wells score as an ordinal test}
\label{section_results_modelling_wells_as_ordinal}
Now we fit the models without dichotomising the Wells score, by simultaneously modelling all three categories. 
For these models, we used weakly informative priors of $ \mu_{3} \sim \text{N}(0, 1) $ for the mean parameters for the Wells test. We used the partial pooling model on the Wells score cutpoint parameters (see section \ref{section_methods_within_study_model}, equation \ref{eq_3_study_specific_accuracy_estimates_ordinal}). 
For the Dirichlet population parameters, we used a weakly informative prior $ \kappa^{[d]} \sim \text{N}_{\ge 0}(0, 50) $. This allows considerable asymmetry in the Dirichlet population vector $ \boldsymbol\alpha_{k}$, as can be seen from the prior predictive check (see figure 1 in section 1.3 in supplementary material 1).
The rest of the priors were the same as those discussed in section \ref{section_results_priors}.

We fit the following models: 
one assuming that ultrasound is a perfect gold standard and conditional independence between all three tests (\textbf{M1}); 
the same model but modelling the conditional dependence between the Well's score and D-Dimer (\textbf{M2}); 
a model assuming ultrasound to be an imperfect gold standard and conditional independence between all three tests (\textbf{M3}); 
and a variation of M3 which modelled the conditional dependence between all three tests (\textbf{M4}). 
\begin{figure}[H]
    \centering
    \includegraphics[width=15cm]{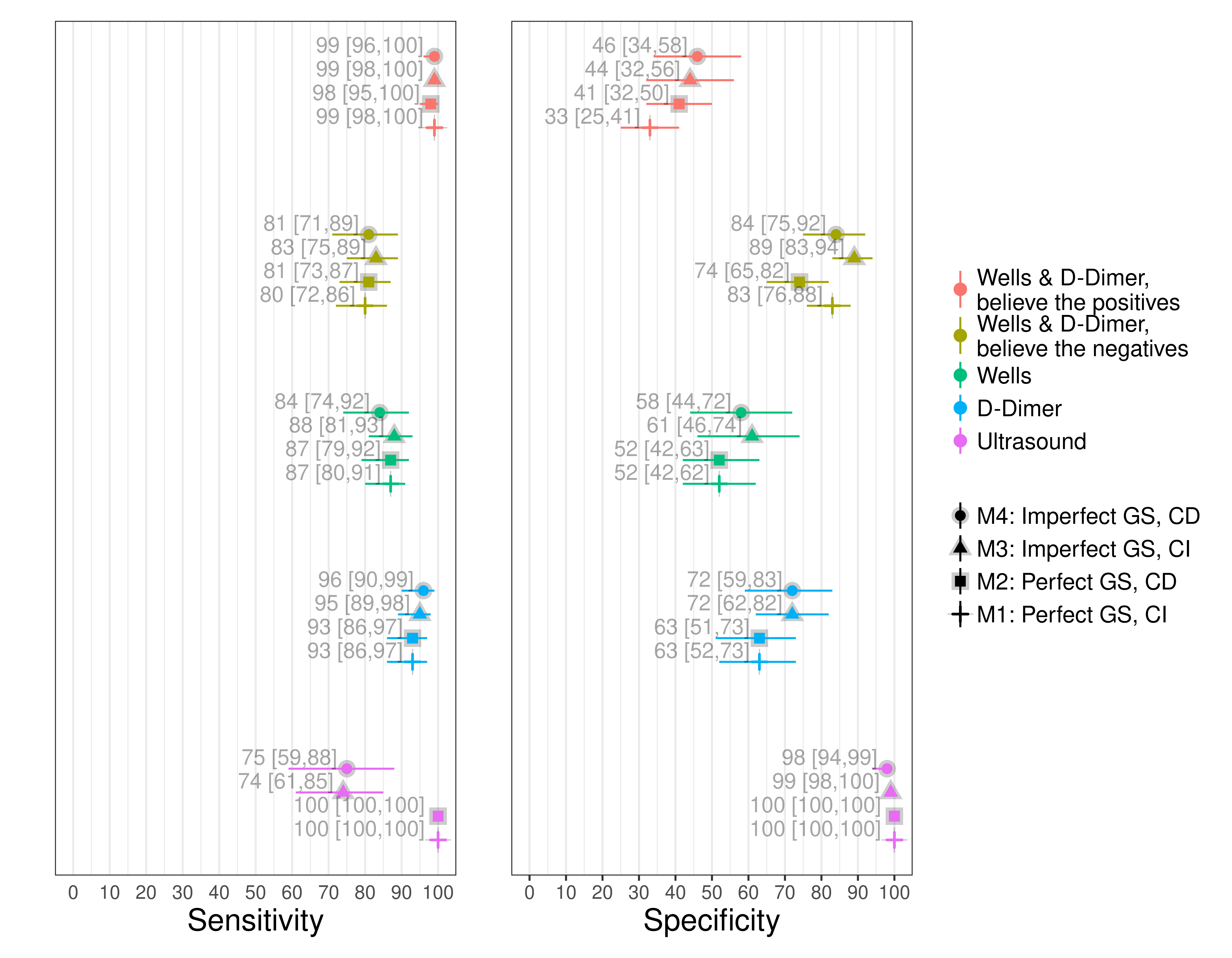}
    \caption{Posterior medians and 95\% posterior intervals for summary sensitivities and specificities, for models 1 - 4. 
    Note: The Wells score summary estimates are dichotomised as low vs moderate + high. CD = Conditional Dependence; CI = Conditional Dependence; GS= Gold Standard.}
    \label{fig3}
\end{figure}

\begin{figure}[H]
    \centering
    \includegraphics[width=15cm]{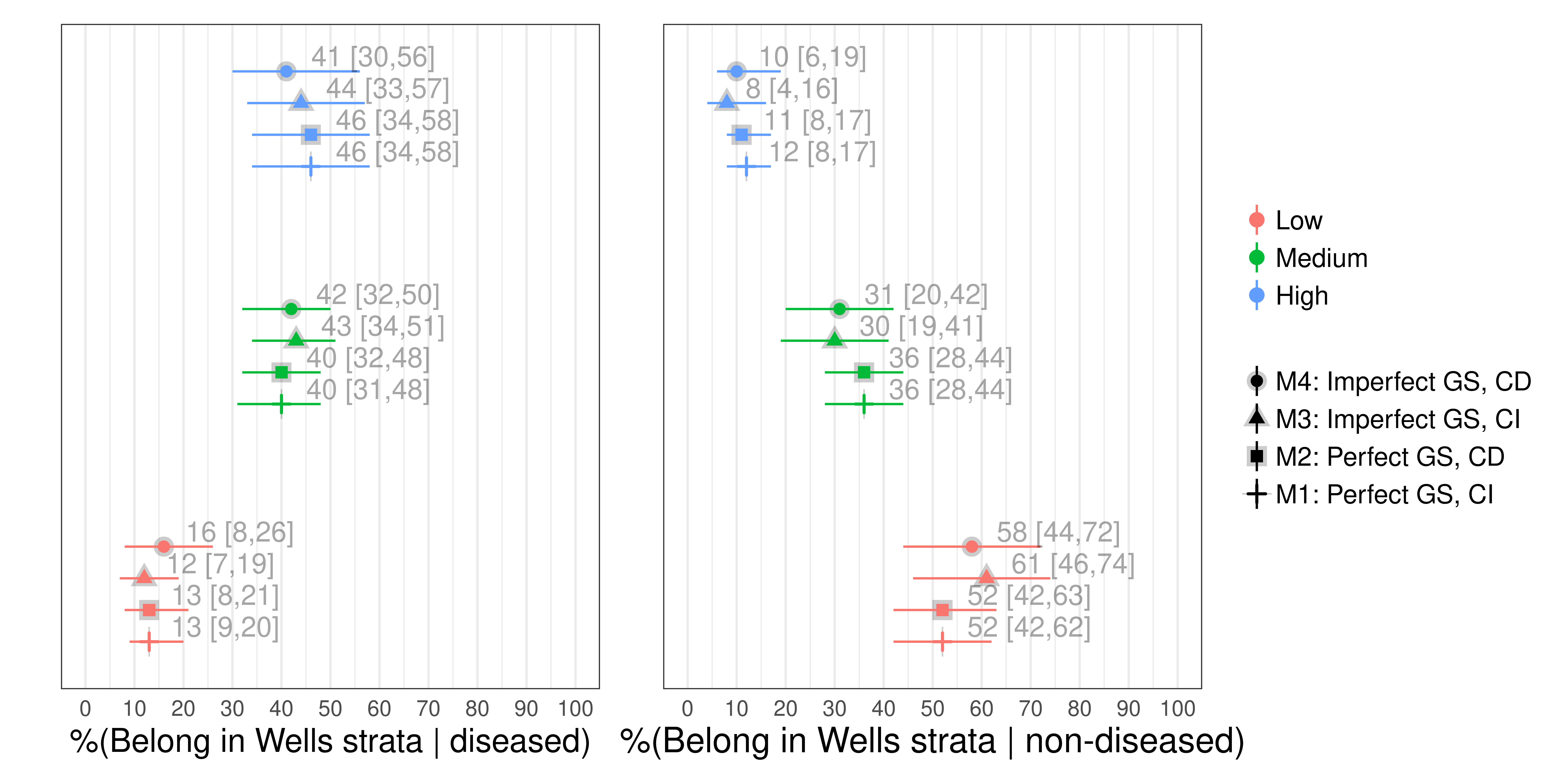}
    \caption{Posterior medians and 95\% posterior intervals for the Well's score stratum, for models 1 - 4. 
    Note: CD = Conditional Dependence; CI = Conditional Dependence; GS= Gold Standard.}
    \label{fig4}
\end{figure}

\begin{figure}[H]
    \centering
    \includegraphics[width=12cm]{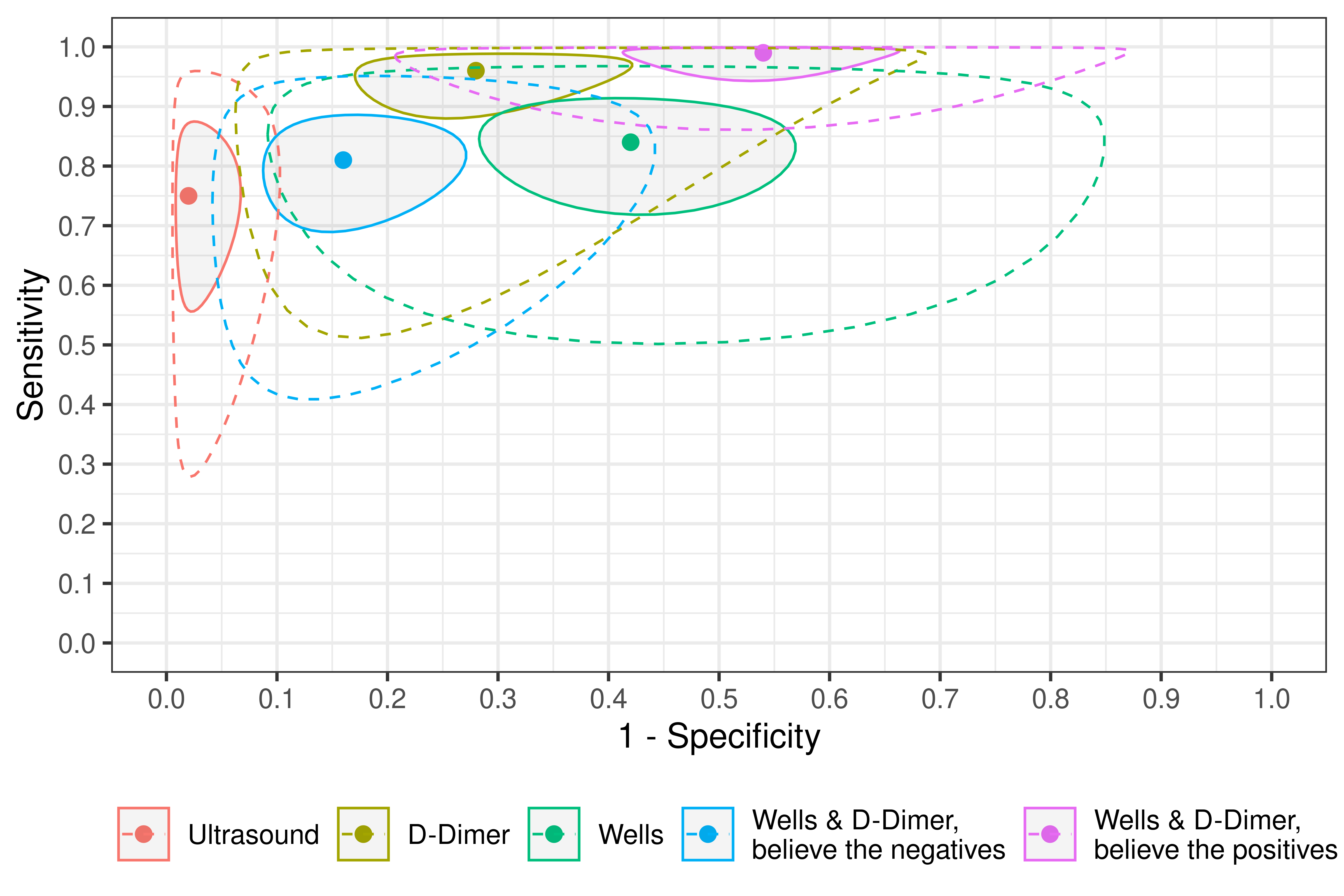}
    \caption{Summary Receiver Operating Characteristic (sROC) plot for M4. Shaded regions represent 95\% posterior regions and regions surrounded by dashed lines represent 95\% prediction regions.
    Note: The Wells score summary estimates are dichotomised as low vs moderate + high}
    \label{fig5}
\end{figure}

The results for the summary sensitivity and specificity estimates for the four models are shown in figure \ref{fig3}, and the results for each of the Wells score strata are shown in figure \ref{fig4}. The estimates for the two models assuming a perfect gold standard (M1 and M2) are within 2\% of those obtained from Novielli et al \supercite{Novielli2013}. The similarity of the results is not surprising, since despite using different models and different link functions (logit vs approximate probit), both models assume that ultrasound is perfect. 

For both the conditional independence (M1) and dependence (M2) models which assumed that ultrasound is a perfect reference test, the results we obtained for the accuracy of the BTP and BTN testing strategies for the Wells \& D-Dimer tests were similar to those obtained by Novielli et al\supercite{Novielli2013}. More specifically, for the  BTP testing strategy, we found that the summary specificity estimates for M1 (33 [25, 41]) were around 8\% lower than M2 (41 [32, 50]). For the BTN strategy, we found that the estimates for M1 (74 [65, 82]) were around 9\% higher (83 [76, 88]) than M2.

When we modelled ultrasound as an imperfect test, the summary estimate for the sensitivity of the Wells test for the model assuming conditional independence (M3 - 88 [81, 93] ) was around 4\% higher than the model which modelled conditional dependence (M4 - 84 [74, 92]), and around 5\% higher for the sensitivity of the Wells \& D-Dimer BTN testing strategy (89 [83, 94] and  84 [75, 92] for M3 and M4, respectively). The other differences between M3 and M4 were 3\% or less (see figure \ref{fig3}).

Whilst assuming conditional dependence, the model assuming ultrasound is perfect estimated the specificity of the Wells score to be around 6\% lower than the conditional dependence model (52 [42, 63] and 58 [44, 72] for M2 and M4, respectively). We also found that the specificity of the D-Dimer was around 9\% lower (63 [51, 73] and 72 [59, 83] for M2 and M4, respectively), that the specificity was around  5\%  lower for the Wells \& D-Dimer BTP testing strategy  (41 [32, 50] and 46 [34, 58] for M2 and M4, respectively), and that the specificity of the BTN testing strategy was around 10\% lower (74 [65, 82] and 84 [75, 92] for M2 and M4, respectively).

The summary receiver operating characteristic plot for M4 is shown in figure \ref{fig5}. The prediction regions suggest that there is substantial between-study heterogeneity for the sensitivity and specificity for most estimates. However, we found relatively narrow prediction regions for the specificity of ultrasound and for the Wells and D-Dimer BTP testing strategy, so we can be more confident in generalising our inferences for these estimates.

The LOO-CV results for all of the models are shown in table \ref{table_loo}. The results suggested that M1 has the poorest fit. Modelling the dependency between the D-Dimer and Wells tests (M2) improved the fit (LOO-IC = 16038.6 and 15819.0 for M1 and M2, respectively). Out of the two models not assuming a perfect gold standard, conditional independence model gave a worse fit than the conditional dependence model (difference in ELPD between M3 and M4 = -31.4, se = 6.4). The two conditional dependence models were the two best fitting models, with the conditional dependence model giving the best fit (difference in ELPD between M2 and M4 = -20.8, se = 6.2). The posterior predictive checks for this model are shown in figure \ref{fig6} (correlation residual plot) and figure 2 in supplementary material 3 (2x2 table count residual plot). Both plots show that the model fits the data well. 

\begin{table}[H]
   \centering
   \begin{threeparttable}
   \caption{Leave-One-Out Cross Validation (LOO-CV) for comparison of model fit for case study 1 dataset}
   \label{tab:test2}
   \begin{tabular}{llll}
   \hline
   Model\tnote{1} & LOO-IC\tnote{2} & $ELPD_{M4} - ELPD_{Mi}$ \tnote{3,4} & $se(ELPD_{M4} - ELPD_{Mi})$ \tnote{4}\\
   \hline
   4 (Imperfect ultrasound + CD) & 15,777.4 &  0     & 0    \\
   2 (Perfect ultrasound + CD) & 15,819.0 & -20.8  & 6.2  \\
   3 (Imperfect ultrasound + CI) & 15,840.1 & -31.4  & 6.4  \\
   1 (Perfect ultrasound + CI) & 16,038.6 & -130.6 & 15.4 \\
   \hline
     \label{table_loo}
   \end{tabular}
   \begin{tablenotes}
     \item[1] Models are ordered from best to worst fitting
     \item[2] LOO-IC = Leave-One-Out Information Criterion; note that LOO-IC is on the deviance scale
     \item[3] ELPD = Estimated Log pointwise Predictive density for a new Dataset
     \item[4] $M{i}$ denotes the $i$th model
     \item CI = Conditional Independence; CD = Conditional Dependence
   \end{tablenotes}
  \end{threeparttable}
\end{table}

\begin{figure}[H]
    \centering
    \includegraphics[width=15cm]{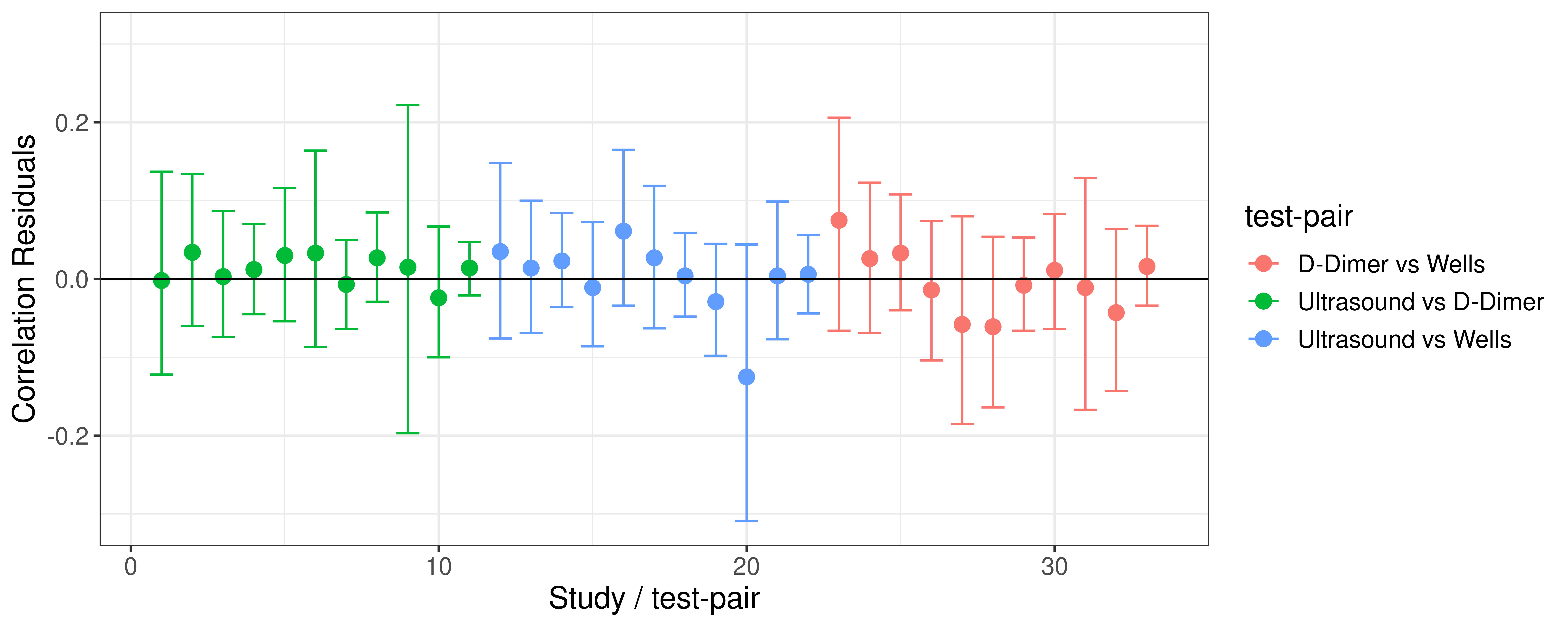}
    \caption{Posterior predictive check for model 4; correlation residual plot}
    \label{fig6}
\end{figure}

\section{Discussion} 
\label{section_discussion}
\subsection{Summary} 
\label{section_discussion_summary}
Our proposed MVP-LC model addresses the novel problem of carrying out meta-analysis of two or more conditionally dependent tests when there is no perfect gold standard, for the case where there are both ordinal and dichotomous test(s) under evaluation, and estimation of joint test accuracy is of interest. 

Using the case study as a demonstrative aid for the model (see section \ref{section_results_application_to_case_study}), we showed why treating ordinal tests as dichotomous in the context of an imperfect gold standard is suboptimal (see section \ref{section_results_a_priori_dichot}).
When we modelled the Wells test as ordinal and treated ultrasound as a perfect gold standard (see section \ref{section_results_modelling_wells_as_ordinal}), the summary estimates from Novielli et al\supercite{Novielli2013} are replicated in our findings.
However, we found that the most complex model - which treated ultrasound as an imperfect gold standard in addition to modelling the conditional dependence between tests - had the best fit to the data. 
For this model, our estimates of test accuracy differed considerably compared  to other models we fit (which gave worse fit to the data) and compared to the results obtained in the analysis conducted by Novielli et al\supercite{Novielli2013}. In particular, we obtained considerably different estimates of specificity for both the D-Dimer and the Wells score tests when used alone, and for the joint specificity of the Wells and D-Dimer BTN testing strategy. However, the large between-study heterogeneity limited the generalisability of our results. 
\subsection{Potential applications} 
\label{section_discussion_potential_implications}
The methods we have developed in this paper have a wide scope of applicability in clinical practice, further than just DVT.
For instance, 
Hamza et at \supercite{Hamza2009} re-analysed a meta-analysis\supercite{aertgeerts_value_2004}, which assessed the accuracy of the CAGE questionnaire \supercite{ewing_detecting_1984} - a 4-category ordinal test used as a screening tool to detect individuals who may be suffering from alcoholism. However, their model assumed a perfect gold standard\supercite{Reitsma2005}.
Our proposed MVP-LC model could be used to more appropriately estimate the accuracy of the CAGE questionnaire, since we would not need to assume that the reference test in each study is perfect. 

The methods could also be used to more appropriately assess joint testing strategies. 
For instance, current UK Health Security Agency guidance \supercite{covid_19_UK_Health_Security_Agency_guidance_wilkinson-brice_may_kanani_powis_2021} states that individuals who have symptoms suggestive of the severe acute respiratory syndrome coronavirus 2 (SARS-CoV-2) and test positive using Lateral flow tests (LFTs) should be considered as positive and require no subsequent testing.
On the other hand, it states that individuals with negative LFTs should be assessed with a polymerase chain reaction (PCR), with only those who also test negative on PCR being considered negative.
Our methods could be used to investigate this joint test accuracy strategy without modelling PCR as an imperfect gold standard, particularly with respect to its sensitivity. 
For depression screening, one potential BTN testing strategy is one in which individuals undertake a very brief 2-item version of the Patient Health Questionnaire (PHQ-2 \supercite{phq_2}) followed by the 9-item version (PHQ-9 \supercite{phq_9}). This was investigated recently by Levis et al\supercite{levis_JAMA_2020}; however, they assumed perfect gold standards, and they only used around half of the available studies, since they discarded those studies which used inferior gold standards.
Our MVP-LC model could be used to analyse these data without assuming a perfect gold standard whilst accounting for differences between reference tests with meta-regression, and using all of the available data. 
Furthermore, we would be able to model the differences in gold standards between studies using meta-regression (see section \ref{section_methods_between_study_model}). 
\subsection{Advantages} 
\label{section_discussion_advantages_over_other_models}
Our proposed Bayesian MVP-LC model addresses some important limitations which are present in models based on TLCMs \supercite{Hui1980, Vacek1985,Chu2009, Menten2013, Dendukuri2012}. 
For example, although TLCMs have fast run times due to being computationally inexpensive, and they can model the conditional dependence between tests\supercite{Wang2017}, an important limitation is that, unlike our proposed MVP-LC model, they cannot appropriately model ordinal tests. For example, if one wishes to simultaneously model ordinal tests whilst modelling conditional dependence, they would first need to dichotomise the data a priori. As we showed in section \ref{section_results_a_priori_dichot}, this is suboptimal in the context of an imperfect gold standard, since the test accuracy and disease prevalence estimates were varied depending on which cutpoint we dichotomise the data at. 
A limitation of TLCMs which have been proposed for meta-analysis\supercite{Chu2009, Menten2013, Dendukuri2012} is that, unless one assumes a complete pooling model between studies, it is not possible estimate summary correlation parameters - parameters which are required to estimate summary-level joint test accuracy. This is due to the fact that, in contrast to our MVP-LC model (which uses the within-study correlations), TLCMs model the conditional dependence using the within-study covariances, making it difficult to construct a partial pooling model for the within-study conditional dependence parameters. These covariances have bounds based on the sensitivity and specificity parameters in each study\supercite{Vacek1985, dendukuri2004}. Therefore, any summary-level covariance parameters obtained would be questionable. 
Our MVP-LC model also has advantages over more advanced models for meta-analysis of test accuracy, such as the model proposed by Sadatsafavi et al\supercite{Sadatsafavi2010}, which is also based on multivariate probit regression and is an extension of the latent trait model \supercite{Qu1996}. Two important limitations of this model - not present in our MVP-LC model - is that it can only model dichotomous data, and it assumes that the within-study correlations are fixed across studies. Furthermore, since our proposed MVP-LC model is an extension of the model for single studies proposed by \supercite{Xu2013}, another benefit over the model by Sadatsafavi et al\supercite{Sadatsafavi2010} is that it can also be used to specify more general correlation structures (by setting certain correlations to zero - see section \ref{section_methods_within_study_model}). The fact that our model is Bayesian means that one can incorporate subject-matter knowledge into the model, as we did for our case study. Furthermore, the Induced Dirichlet partial pooling model \supercite{Betancourt2017_hmc_intro} (see section \ref{section_methods_between_study_model_cutpoints}) and supplementary material 4) for the ordinal tests makes it possible to specify priors for ordinal tests and obtain summary estimates.
\subsection{Limitations}
\label{section_discussion_limitations}

When applying the model to our case study dataset (see section \ref{section_case_study_dataset_motivating_example}), we used available subject-matter knowledge\supercite{Goodacre2005} to construct informative prior distributions for the gold standard test (ultrasound), and weakly informative priors for other parameters (see section\ref{section_results_priors} and supplementary material 1).  Attempts to conduct sensitivity analysis using more diffuse priors led to diagnostic errors. This is likely due to the fact that Stan is quite sensitive at detecting non-identifiabilities in the posterior distributions\supercite{Carpenter2017}, and non-identifiability is more likely to occur with less informative priors, particularly for latent class models due to the large number of parameters relative to the data. Another limitation of our case study analysis is that, although our model can easily incorporate meta-regression coefficients (see supplementary material 1), the case study dataset did not contain any study-level covariates, since primary studies did not report sufficient data. In an ideal world where such data were available, a more principled analysis could be carried out by using a meta-regression covariate for the proportion of patients who have proximal versus distal DVT, which would have enabled us to model the variation of ultrasound sensitivity that exists between the two DVT groups in Novielli et al's\supercite{Novielli2013} data.

A limitation of our model, which is present across all imperfect gold standard methods based on latent class models (including TLCMs), is that full cross-classification tables (i.e. the full distribution of test results) are required for each study. This is a potential barrier to the uptake of our proposed MVP-LC model, as this data is frequently not reported for studies evaluating 3 or more tests and/or studies assessing ordinal tests.
One way in which we could have assessed the general performance of our MVP-LC model is by running a simulation study \supercite{morris_using_2019}. A simulation study comparing our proposed MVP-LC model to other models would also be very useful. However, it is important to note that, at the time of writing, no other models have been proposed to simultaneously meta-analyse both dichotomous and ordinal tests without assuming a perfect gold standard. That being said, a simulation study would still be useful, since we could compare the performance of our model to other proposed models which do assume a perfect gold standard (e.g., Novielli et al \supercite{Novielli2013}) under a variety of different scenarios.

Although our proposed MVP-LC model offers considerable benefits in comparison to the more commonly used TLCM models \supercite{Chu2009, Menten2013, Dendukuri2012, Wang2017} (see section \ref{section_discussion_advantages_over_other_models}), we found that our proposed model was considerably less time efficient than TLCM models. Although this was not prohibitive for the case study used in this paper\ref{section_case_study_dataset_motivating_example}, our MVP-LC model may be intractable for larger sample sizes. Speeding up models based on augmented continuous data, such as our MVP-LC model, is an active area of research \supercite{variational_blei, Duan_cda, duan2020transport, margossian2020hamiltonian_laplace, dinh2017density_normalizingflow, papamakarios2018masked_normalizingflow, rezende2016variational_normalizingflow, Dhaka_Akash_2020}. An important area for future research would be to apply the models developed in this paper using these more efficient algorithms, which would make our proposed MVP-LC model more suitable for general use with larger meta-analyses, and it would also make it easier to conduct more meaningful simulation studies.

\subsection{Future work}
\label{section_discussion_future_work}

Models for the meta-analysis of test accuracy which can incorporate patient-level covariates - otherwise known as individual patient data (IPD) - have been proposed \supercite{riley_ipd_2008}, but only for dichotomous data and they assume a perfect gold standard  \supercite{riley_ipd_2008}.  Modelling IPD can lead to results which are more applicable to clinical practice as they can more easily be applied to patients when there is between-study heterogeneity, rather than only providing summary estimates which relate to some "average" patient. Extending our model to incorporate IPD would be relatively straightforward, since our model uses the patient-level data (as reconstructed from the reported contingency tables) as opposed to aggregated data for each study. 
It is straightforward to extend our model to the case where not all studies are assessing the same number of tests, using direct comparisons only. This could be further extended to allow indirect comparisons (network meta-analysis [NMA]), by assuming tests are missing at random (MAR) \supercite{Rubin1976}, and extending the between-study model described in section \ref{section_methods_between_study_model} to an arm-based network-meta analysis model\supercite{Ma2018, Nyaga2018}.
Another straightforward modelling extension would be to incorporate data from ordinal tests which have missing data for some categories. 

Our model could also be extended to synthesize data from ordinal tests for the case where some (or all) studies do not report data for every cutpoint - which is common in research. One could formulate such a 'missing cutpoint' version of our MVP-LC model by extending the partial pooling between-study cutpoint model (see section \ref{section_methods_between_study_model_cutpoints} and supplementary material 4), and viewing the cutpoints as MAR.
Another possible 'missing cutpoint' model could be constructed by modelling the cutpoint parameters as the same in the diseased and non-diseased classes, and assume that they are fixed between studies by using a no pooling model. Then, as opposed to our MVP-LC model, in which the within-study variances are set to 1 to ensure parameter identifiability (see section \ref{section_methods_within_study_model}), the no pooling cutpoint model would allow us to introduce within-study variance parameters and model them using a partial pooling model without encountering significant identifiability issues. These within-study parameters could be set to vary between the two latent classes, which would result in a smooth, non-symmetric receiver operating characteristic (ROC) curve.
Another possible 'missing cutpoint' approach would be one based on the model proposed by Dukic et al\supercite{dukic}, which assumes a perfect gold standard. This model also results in a smooth, non-symmetric ROC curve, since it assumes that the cutpoints vary between studies and are the same in the diseased and non-diseased class. However, it would be more parsimonious since it assumes that the sensitivity is some location-scale change of the false positive rate. 

For the case where studies report thresholds at explicit numerical cutpoints (as is sometimes reported for continuous tests, such as biomarkers), some 'missing threshold' methods which assume a perfect gold standard have been proposed\supercite{Jones2019, Steinhauser2016}. Rather than modelling the cutpoints as parameters, these methods assume that the cutpoints are constants, equal to the value of the numerical cutpoint, and they estimate separate location and scale parameters in each study and disease class. Our MVP-LC model could be extended to achieve this without assuming a gold standard. 
An important area for future research would be to construct other models which can be used for the same purposes as our proposed MVP-LC model. For instance, a multivariate logistic regression model could be constructed by using the Bayesian multivariate logistic distribution proposed by O' Brien et al\supercite{multivariate_logistic_obrien}. Such a model would use logistic link functions as opposed to probit (or approximate probit) links like our MVP-LC model, which are more numerically stable than probit and may give better fit to some datasets. Another multivariate regression approach would be to use copulas \supercite{copula_mvp_winkelmann, copula_discretechoice_Eichler, copula_bivariate_normal_Meyer}. Besides multivariate regression based on augmented data, another approach to modelling conditionally dependent ordinal diagnostic tests without assuming a perfect gold standard is log-linear models \supercite{Xu2013}. These models can account for higher-order correlations \supercite{Xu2013}. However, this requires estimation of additional parameters, so it is likely to introduce identifiability issues. Similarly to the multivariate probit models utilised in this paper, it may be possible to extend these models to meta-analyse multiple, imperfect diagnostic tests with multiple cutpoints.


\newpage
\section*{Highlights}
\textbf{What is already known?}: Standard, well-established methods exist for the synthesising estimates (i.e., conducting a meta-analysis) of test accuracy. These methods estimate test accuracy by comparing test results to some test which is assumed to be perfect - which is referred to as a 'gold standard' test. However, in clinical practice, these tests are often imperfect, which can cause estimates of the tests being evaluated to be biased and potentially lead to the wrong test being used in clinical practice. Meta-analytic methods, which do not assume a gold standard, have previously been proposed, but only for dichotomous tests. 

\textbf{What is new?}: We developed a model which allows one to simultaneously meta-analyse ordinal and dichotomous tests without assuming a gold standard. The model also allows one to obtain summary estimates of the accuracy of two tests used in combination (i.e. joint test accuracy). 

\textbf{Potential impact for Research Synthesis Methods readers outside the authors field}: The methods are widely applicable. For instance, psychometric measures and radiologic tests are typically ordinal, and the studies assessing these tests often do not use a gold standard; hence, applying standard models to these datasets may lead to misleading conclusions. The methods we proposed may lead to less biased accuracy estimates, and hence potentially a better understanding of which tests, and test combinations, should be used for these conditions. 

\section*{Acknowledgements}
The authors would like to thank Elpida Vounzoulaki for proofreading the manuscript. The authors would also like to thank various members of the Stan community forums (see https://discourse.mc-stan.org/) including Ben Goodrich, Michael Betancourt, Stephen Martin, Staffan Betnér, Martin Modrák, Niko Huurre, Bob Carpenter and Aki Vehtari and for providing functions which were utilised in the models and for useful discussions. 

Funding: The work was carried out whilst EC was funded by a National Institute for Health Research (NIHR) Complex Reviews Support Unit (project number 14/178/29) and by a National Institute for Health Research Systematic Review Fellowship (project number RM-SR-2017-09-023). The views and opinions expressed herein are those of the authors and do not necessarily reflect those of the NIHR, NHS or the Department of Health. The NIHR had no role in the design of the study and collection, analysis, and interpretation of data and in writing the manuscript. This project is funded by the NIHR Applied Research Collaboration East Midlands (ARC EM). The views expressed are those of the authors and not necessarily those of the NIHR or the Department of Health and Social Care. 

\section*{Data availability statement}
The Data, R and Stan code to reproduce the results and figures from section \ref{section_results_application_to_case_study} is available on Github at: \\
\href{https://github.com/CerulloE1996/dta-ma-mvp-1}{https://github.com/CerulloE1996/dta-ma-mvp-1}.  
\printbibliography
\setcounter{figure}{0} 
\clearpage 
\clearpage

\end{document}